\documentclass[%
 reprint,
 amsmath,amssymb,
 aps,
 prd,
]{revtex4-2}

\usepackage{graphicx}
\usepackage{dcolumn}
\usepackage{bm}
\usepackage{color}
\usepackage{mathtools}

\DeclarePairedDelimiter{\ceil}{\lceil}{\rceil}

\def\l{\ell}
\def\a{\alpha}
\def\b{\beta}

\newcommand{\barz}{{\bar z}}
\newcommand{\dd}{{\Delta}}

\def\ldef{\mathrel{\mathop:}=}

\newcommand{\x}{{\rho}}
\newcommand{\y}{{\bar\rho}}

\newmuskip\pFqskip
\mathchardef\pFcomma=\mathcode`, 

\newcommand*\pFq[5]{%
  \begingroup
  \begingroup\lccode`~=`,
    \lowercase{\endgroup\def~}{\pFcomma\mkern\pFqskip}%
  \mathcode`,=\string"8000
  {}_{#1}F_{#2}\biggl[\genfrac..{0pt}{}{#3}{#4};#5\biggr]%
  \endgroup
}

\def\mysec#1{\paragraph*{#1 ---}}

\begin{document}

\title{Exact 3D Conformal Blocks from Fractional Calculus}

\author{Chaoming Song}
\email{c.song@miami.edu}
\affiliation{%
Department of Physics, University of Miami, Coral Gables, Florida 33142, USA}%

\begin{abstract}
We uncover a striking connection between conformal blocks and fractional calculus. By employing a modified form of half-derivates, we derived explicitly the exact form of the three-dimensional conformal block, expressed as the product of two hypergeometric \({}_4 F_3\) functions. This result provides a rigorous proof of Hogervorst’s formula, conjectured nearly a decade ago. Furthermore, we demonstrate its implications for the conformal bootstrap, potentially leading to new analytical techniques and numerical tools that deepen our understanding of conformal field theory.
\end{abstract}

\maketitle
\newpage

\mysec{Introduction}
The global conformal block is a fundamental component of the modern conformal bootstrap, playing a crucial role in solving and constraining conformal field theories~\cite{el2012solving,poland2016conformal,simmons2017conformal,poland2019conformal,rychkov2024new}. Considerable effort has been devoted to deriving and understanding these blocks, which exhibit remarkable mathematical structures. Notably, they have been shown to be solutions of integrable systems~\cite{PhysRevLett.117.071602}. While closed-form expressions for conformal blocks are well established in two and four dimensions (2D and 4D), a similarly compact form in three dimensions (3D) has remained elusive~\cite{dolan2004conformal}. 

Since the conformal group in higher dimensions naturally contains lower-dimensional ones, a \((d+1)\)-dimensional conformal block can be decomposed into a sum of \(d\)-dimensional blocks. Surprisingly, Hogervorst demonstrated that the coefficients in this decomposition follow a remarkably simple rational form~\cite{hogervorst2016dimensional}. Though this observation has been verified manually for low-order terms and is widely accepted, its exactness remains a conjecture~\cite{pal2023twist}. Moreover, Hogervorst's formula has gained increasing importance in subsequent developments. Resolving this question has significant implications for both analytical results~\cite{pal2023twist,caron2023leading} and numerical bootstrap methods~\cite{simmons2017lightcone}.

In this Letter, we uncover a surprising connection between conformal blocks and fractional calculus. Previous attempts to simplify conformal blocks have used differential operators of order \( \nu \coloneqq (d-2)/2 \)~\cite{simmons2014projectors}. However, these approaches apply only for integer \( \nu \), corresponding to even dimensions, and do not involve genuinely fractional orders. In contrast, we demonstrate that 1D and 2D conformal blocks can be expressed as power functions of the radial coordinate after a transformation involving a half-derivative operator \( \mathcal{T}^{1/2} \). This insight allows us to explicitly solve the 3D conformal Casimir equation as a product of two hypergeometric \({}_4 F_3\) functions, leading to a proof of Hogervorst’s formula. Beyond providing a compact representation for the 3D conformal block, our results establish a new paradigm for understanding conformal block expansions through fractional calculus, with potential applications extending beyond the bootstrap program.

\mysec{Fractional Derivative}
We first review the concept of fractional derivatives, beginning with the Riemann-Liouville fractional derivative:
\begin{equation}
    \frac{d^{-\delta}}{d x^{-\delta}} f(x) \ldef (I^\delta f)(x) \ldef \frac{1}{\Gamma(\delta)} \int_0^x f(t) (t-x)^{\delta-1} dt,\notag
\end{equation}
for $\delta > 0$. Otherwise, the formula can be analytically continued by applying a finite number of ordinary derivatives and shifting $\delta$, i.e., $\frac{d^{\delta }}{d x^\delta} \ldef \frac{d^{\ceil{\delta} }}{d x^{\ceil{\delta}}} I^{\ceil{\delta}-\delta}$ \cite{miller1993introduction}. The Riemann-Liouville fractional derivative is consistent with the regular derivative when $\delta$ is an integer. Additionally, it satisfies the semigroup property of differentiation,
$\frac{d^{\alpha }}{d x^\alpha} \frac{d^{\beta }}{d x^\beta}  = \frac{d^{\alpha+\beta }}{d x^{\alpha+\beta}}$. Based on this, we introduce the operator
\begin{align} \label{eq:T}
     T_{x}^{\delta} \ldef \frac{\mathrm{d}^{-\delta}}{\mathrm{d}(-1/x)^{-\delta}},
\end{align}
which applies the transformation \( x \rightarrow -1/x \) and \( f(x) \rightarrow f(-1/x) \) to the Riemann-Liouville integral. Equivalently, we have 
\begin{align}\label{eq:T1}
(T^\delta_x f)(x) &= \frac{x^{1-\delta}}{\Gamma(\delta)} \int_0^x f(t) (x-t)^{\delta-1}  t^{-\delta-1} d t
\notag\\&= \frac{x^{-\delta}}{\Gamma(\delta)} \int_0^1 f(xt)(1-t)^{\delta-1}  t ^{-\delta-1} d t,
\end{align}
provided the integral converges; otherwise, an appropriate analytic continuation is used. The operator $T_x^{\delta}$ is also a special case of the Erdélyi-Kober operator \cite{sneddon2006use},
\[
(K^{\alpha,\beta} f)(x)  \ldef \frac{x^\beta}{\Gamma(\beta)}\int_0^x f(t)(x-t)^{\alpha-1}  t^{\beta-1}  d t
\]
with \( T^\delta_x \) being equivalent to \( x K^{\delta,-\delta} \). It is straightforward to verify that 
\begin{equation}\label{eq:power}
T^{\delta} x^h = \frac{\Gamma(h-\delta)}{\Gamma(h)} x^{h-\delta},
\end{equation}
lowering the power exponent by the order $\delta$. Additionally, it satisfies the semigroup property,
\begin{equation}\label{eq:com}
    T_x^{\alpha}T_x^{\beta} = T_x^{\beta}T_x^{\alpha} = T_x^{\alpha+\beta}.
\end{equation}
Thus, the operator $T^{\delta}_x$ can be viewed as a variation of fractional derivatives.

The fractional derivative $T^{\delta}_x$ enjoys several notable algebraic properties. Two identities particularly useful later are
\begin{equation}\label{eq:xD}
    T_x^\delta\, x D_x\, T_x^{-\delta} = xD_x, 
    \qquad 
    T_x^\delta\, D_x\, T_x^{-\delta} = D_x+\delta,
\end{equation}
where $D_x \ldef x\partial_x$ is the dilation operator. 
The first relation follows from Eq.~\eqref{eq:T}, which gives $T^{-1}_x=\partial_{-1/x}=xD_x$, 
while the second follows from the commutation relation~\eqref{eq:com} 
together with the identity $D_x=-D_{-1/x}$.

\mysec{Conformal Blocks and Crossing Symmetry}
In conformal field theory, a central object is the four-point correlator of identical scalar operators
\(
\langle \phi(x_1)\phi(x_2)\phi(x_3)\phi(x_4)\rangle,
\)
here $\phi$ has scaling dimension $\Delta_\phi$. This correlator depends only on the conformally invariant cross-ratios
\(
z,\bar z.
\)
Its decomposition into conformal blocks $G_{\Delta,\ell}(z,\bar z)$ encodes the contributions of intermediate primary operators with scaling dimension $\Delta$ and spin $\ell$. In the $s$-channel this takes the operator product expansion (OPE)
\[
C(z,\bar z) \ldef 1+\sum_{\Delta,\ell} c_{\Delta,\ell}\, G_{\Delta,\ell}(z,\bar z),
\]
with positive OPE coefficients $c_{\Delta,\ell}$ in unitary theories. Crossing symmetry requires equality with the $t$-channel decomposition, imposing the constraint
\begin{equation}
  C(z,\bar z) \;=\; \left(\frac{z \bar z}{(1-z)(1-\bar z)}\right)^{\Delta_\phi} C(1-z,1-\bar z).
\label{eq:crossing0}
\end{equation}

For certain spacetime dimensions, compact formulas for the conformal blocks $G_{\Delta,\ell}(z,\bar z)$ are explicitly known. In $d=1$ there is only a single cross-ratio $z$ and no spin dependence, and the block reduces to
\begin{equation}
    k_{h}(z) \ldef 4^{1/2-h}\,\frac{\Gamma(h)}{\Gamma(h-1/2)}\, z^h \, {}_2 F_1(h,h,2h;z).
\end{equation}
In $d=2$ and $d=4$, closed-form expressions involving $k_h(z)$ are available~\cite{dolan2011conformal}, indicating that the 1D block acts as a fundamental building block for higher-dimensional cases. By contrast, in $d=3$ no compact analytic form has been known, aside from series expansions or recursion relations. This gap has long been a central obstacle for both analytic and numerical bootstrap studies, and it is precisely this problem that our approach resolves.

Our main application of fractional derivatives $T_x^\delta$ to conformal blocks will focus on $\delta = \pm 1/2$, where $T^{-1/2}_x$ represents the half-derivative under the variable change $x\to -1/x$, and $T^{1/2}_x$ denotes its inverse. Specifically, we have (see Supplemental Material~\cite{SM} Sec.~A)
\begin{subequations}
\begin{align}
&(T^{-1/2}_x f)(x) = \frac{x^{1/2}}{\sqrt\pi} \int_0^x f'(t) \sqrt{\frac{t}{x-t}}\, dt, \\
&(T^{1/2}_x f)(x)' = \frac{x^{-3/2}}{\sqrt\pi} \int_0^x f'(t) \sqrt{\frac{t}{x-t}}\, dt,
\end{align}
\label{eq:Thalf}
\end{subequations}
which converge provided $f(t)$ scales faster than $t^{-1/2}$ as $t\to 0$.

A striking simplification emerges from these operators. Using Eq.~\eqref{eq:power}, one verifies that
\(
T_z^{1/2} k_h(z) = \Bigl(\tfrac{z}{4}\Bigr)^{h-1/2}\, {}_2F_1(h,h-1/2,2h;z).
\)
Applying the quadratic map \(z = 4\x/(1+\x)^2\), \(\bar{z} = 4\y/(1+\y)^2\), this further reduces to
\begin{align}\label{eq:h}
T^{1/2}_z k_h(z) = \x^{h-1/2}, \quad
T^{-1/2}_z \x^{h} = k_{h+1/2}(z),
\end{align}
where $\x$ and $\y$ are the radial coordinates~\cite{hogervorst2013radial} (see Supplemental Material~\cite{SM} Sec.~B). Equation~\eqref{eq:h} demonstrates that the half-derivative trivializes the 1D block into a simple monomial, revealing a deep and unexpected connection between fractional calculus and conformal blocks.

\mysec {Conformal Casimir Equation}
Conformal blocks are distinguished by the fact that they are eigenfunctions of the quadratic Casimir operator of the conformal group~\cite{dolan2011conformal}. This property provides a differential equation that uniquely characterizes the blocks for given scaling dimension $\Delta$ and spin $\ell$: 
\begin{equation}\label{eq:casimir}
    \mathcal{D} G_{\Delta,\ell}(z,\bar z) = \tfrac{1}{2} C_{\Delta,\ell}\, G_{\Delta,\ell}(z,\bar z).
\end{equation}
Here, the eigenvalue is
\(
C_{\Delta,\ell} \ldef \Delta(\Delta-d)+\ell(\ell+d-2) = 2(\alpha^2+\beta^2 - \nu(\nu+1))-1,
\)
with $\alpha \ldef (\Delta + \ell - 1)/2$ and $\beta \ldef (\Delta-(\ell+d-1))/2$. The quadratic Casimir operator takes the form
\[
\mathcal{D} \ldef \mathcal{D}_z + \mathcal{D}_{\bar z} + 2\nu \frac{z\bar z}{z-\bar z}\left((1-z)\partial_z - (1-\bar z)\partial_{\bar z}\right),
\]
with $\mathcal{D}_z \ldef z^2(1-z)\partial_z^2 - z^2\partial_z = (D_z - 1 - zD_z)D_z$. For most of this paper, we restrict our discussion to identical external operators unless explicitly stated otherwise.

Rather than solving Eq.~\eqref{eq:casimir} directly, we introduce an alternate representation using the fractional derivatives $T^{\pm1/2}$:
\begin{equation}\label{eq:casimir2}
    \tilde {\mathcal{D} }\tilde G_{\dd,\l} = \frac{1}{2} C_{\dd, \l}   \tilde G_{\dd,\l},
\end{equation}
where \( \tilde {\mathcal{D} } \ldef \mathcal{T}^{1/2} \mathcal{D} \mathcal{T}^{-1/2} \) with \(\mathcal{T}^{\delta} \ldef T^{\delta}_z T^{\delta}_\barz\), and
\begin{equation}
    \tilde G_{\dd,\l} = \mathcal{T}^{1/2} G_{\dd,\l}, \quad  G_{\dd,\l} = \mathcal{T}^{-1/2} \tilde  G_{\dd,\l},\notag
\end{equation}
which establishes a transformation between the modified conformal block \( \tilde G_{\dd,\l} \) and the conventional conformal block \( G_{\dd,\l} \).

We will see that the Casimir equation~\eqref{eq:casimir2} in this new representation leads to notable simplifications. To illustrate this, consider the two-dimensional case (\(\nu = 0\)), where \( \tilde {\mathcal{D} } =  \tilde {\mathcal{D} }_z +  \tilde {\mathcal{D} }_\barz\), with \( \tilde {\mathcal{D}}_z \ldef T_z^{1/2} \mathcal{D}_z T_z^{-1/2} \). By applying Eq.~\eqref{eq:xD}, we find
\(
\tilde{\mathcal{D}}_z = (D_z - \frac{1}{2} - zD_z)(D_z + \frac{1}{2}),
\)
or equivalently,
\begin{equation}\label{eq:Dx}
\tilde {\mathcal{D}}_z  = D_\x^2 - 1/4, \notag
\end{equation}
where we use the identities \(D_z = \frac{1+\x}{1-\x} D_\x\) and \(zD_z = \frac{4\x}{1-\x^2} D_\x\). Consequently, the Casimir equation~\eqref{eq:casimir2} simplifies to
\begin{equation}\label{eq:casimir2d}
\left(D_\rho^2 + D_{\bar\rho}^2 \right) \tilde G^{(2d)}_{\dd,\l}(\rho, \bar \rho) = \left(\a^2+\b^2\right)\tilde G^{(2d)}_{\dd,\l}(\rho,\bar\rho),
\end{equation}
which has the solution
\begin{equation}\label{eq:2d}
   \tilde G_{\dd,\l}^{(2d)}(\x,\y) =  \x^{\a} \y^{\b} + \x^{\b} \y^{\a}.  
\end{equation}
Utilizing Eq.~\eqref{eq:h}, we obtain the well-known form for the 2D conformal block
\[
G_{\dd,\l}^{(2d)}(\x,\y) = k_{\a+1/2}(z)k_{\b+1/2}(\barz) + k_{\b+1/2}(z)k_{\a+1/2}(\barz).
\]

Now, we consider the case in general dimensions. To handle the fractional factor \(z-\barz\) in Eq.~\eqref{eq:casimir}, we apply the operator \((D_z-1)(D_{\bar z}-1) (z-\bar{z})\) on both sides before transforming into the new representation, leading to
\begin{widetext}
\begin{equation}\label{eq:master}
\begin{split}
\x  \left( (  D_\y - 1/2 )   D_\x \left( D_\x^2 + (D_\y-\nu)^2 - \a^2-\b^2\right)   - \x\y ( D_\x +1/2)  D_\y \left( (D_\x+\nu)^2 + D_\y^2 - \a^2-\b^2 \right) \right) \tilde G_{\dd,\l} \\ = 
\y  \left( (  D_\x - 1/2 )   D_\y \left( (D_\x-\nu)^2 + D_\y^2 - \a^2-\b^2\right)   - \x\y ( D_\y +1/2)  D_\x \left( D_\x^2 + (D_\y+\nu)^2 - \a^2-\b^2 \right) \right) \tilde G_{\dd,\l}, 
\end{split}
\end{equation}
\end{widetext}
which is the conformal Casimir equation in our representation and naturally reduces to Eq.~\eqref{eq:casimir2d} for  the 2D case (\(\nu = 0\)). 

\mysec{3D Conformal Block}
To solve for general \(\nu\), we use the Frobenius method, expanding the solution in terms of a power series $\tilde G_{\dd,\l}(\x,\y) = \x^\frac{\dd-\l-1}{2} \y^\frac{\dd+\l-1}{2} \sum_{n,m\geq 0} A_{n,m} \x^{n+m} \y^{n-m}$,
where the coefficient \( A_{n,m} = A_{n,\l-m} \) is symmetric to ensure that \( G_{\dd, \l} \) remains symmetric under the exchange of \(\x\) and \(\y\).  Substituting into Eq.~\eqref{eq:master} yields a recursive relationship:
\begin{align}
&f_1(n,m)A_{n,m} - f_2(n, m) A_{n-1,m} \notag \\ 
&- f_3(n,m) A_{n,m-1} + f_4(n, m) A_{n-1,m-1} = 0,
\label{eq:rec}
\end{align}
with the boundary condition that \( A_{n,m} = 0 \) for \( n < 0 \), \( m < 0 \), or \( m > \l \), and \( A_{0,0} = A_{0,\l} = 1 \). The explicit forms of \( f_i(n,m) \) (for \( i = 1,2,3,4 \)) are rather involved (see Supplemental Material~\cite{SM} Sec.~C). 

Nevertheless, the lowest-order solutions have compact forms for the scalar block for \(\ell = 0\), as:
\begin{align}
    \tilde G_{\dd, 0} = (\x\y)^{\frac{\dd-1}{2}} \pFq{3}{2}{(\dd-1)/2, \dd-1, \nu}{(1+\dd)/2, \dd-\nu}{\x\y}, \notag
\end{align}
or equivalently,
\begin{equation}
    G_{\dd,0} = \sum_{n\geq 0} \frac{(\nu)_n\left(\frac{\dd-1}{2}\right)_n (\dd-1)_n}{n! (\dd-\nu)_n\left(\frac{\dd+1}{2}\right)_n } k_{n+ \dd/2}(z) k_{n+ \dd/2} (\bar{z}). \notag
\end{equation}
This result has not been previously reported. By verifying the first few terms, we find it consistent with the established result
\(
G_{\dd,0} = u^{\dd/2}\sum_{m,n=0}^\infty \frac{\left[(\dd/2)_n(\dd/2)_{n+m})\right]^2}{n!m!\left(\dd-\nu\right)_n(\dd)_{2n+m}}u^n(1-v)^m,
\)
where \( u \ldef z\barz \) and \( v \ldef (1-z)(1-\barz) \) (see also numerical validation in Supplemental Material~\cite{SM} Sec.~F). However, a direct proof of their equivalence appears highly nontrivial. Moreover, our new result replaces the double summation with a single summation at the cost of introducing the hypergeometric function. 

For general spin \( \l \), in the 2D case (\(\nu = 0\)), one readily verifies in Eq.~\eqref{eq:rec} that the coefficients \( A_{n,m} \) are nonzero only for \( n=0 \) and \( m=0, \l \). Thus, setting \( A_{n,m} = \delta_{n,0} ( \delta_{0,m} + \delta_{0,\l-m}) \) leads to Eq.~\eqref{eq:2d}. Interestingly, we also find a simplified form of \( f_i \) for the 3D case (\(\nu = 1/2\)), where Eq.~\eqref{eq:rec} reduces to \(f_{n,m}A_{n,m} - f_{n-1/2, m} A_{n-1,m} - f_{n,m-1/2} A_{n,m-1} + f_{n-1/2, m-1/2} A_{n-1,m-1} = 0,\)
with \( f_{n,m} = g_n - h_m \), where
\(
g_n \ldef n(n+\a)(n+\b)(n+\a+\b)\) and~\(\quad h_m \ldef m(m-\a)(m+\b)(m-\a+\b).
\)
This results in the symmetric coefficient
\begin{align}
A_{n,m} = &\frac{(1/2)_n(1/2 +\a)_n (\b+1/2)_n (1/2 + \a +\b)_n}{n!(1+\a)_n (1+\b)_n (1+\a+\b)_n }  \notag \\ 
\times &\frac{(1/2)_m(1/2 -\a)_m (\b+1/2)_m (1/2 -\a+\b)_m}{m!(1-\a)_m (1+\b)_m (1-\a+\b)_m }.\notag
\label{eq:A}
\end{align}   
which leads to the following compact form for the 3D conformal block
\begin{widetext}
\begin{equation}\label{eq:G}
\tilde G_{\dd,\l}^{(3d)}(\x,\y)  =  \x^{\frac{\dd-1-\l}{2}} \y^{\frac{\dd-1+\l}{2}}\pFq{4}{3}{\a+1/2, \b+1/2,\a+\b+1/2, 1/2}{1+\a,1+\b,1+\a+\b}{\x\y} \pFq{4}{3}{1/2-\a, \b+1/2,1/2-\a+\b, 1/2}{1-\a,1+\b,1-\a+\b}{\x/\y}, 
\end{equation}
\end{widetext}
as a product of two ${}_4F_3$ functions, where \( \alpha \ldef (\dd + \ell-1)/2 \) and \( \beta \ldef  (\dd-\ell-2)/2 \). Note that the $1/2-\alpha+\beta = -\ell$ is a negative integer, which implies that the second ${}4F_3$ function truncates to a polynomial. The first ${}4F_3$ function also exhibits good analytic behavior. For example, it converges as $\rho\bar\rho \to 1$. As a result, the transformed block $\tilde G_{\Delta,\ell}$ converges faster and more smoothly than the standard block $G_{\Delta,\ell}$, which typically develops logarithmic divergences at $z, \barz \to 1$. 

Moreover, applying the inverse transform $\mathcal{T}^{-1/2}$ yields the series expansion in terms of 2D conformal blocks $G^{(3d)}_{\dd,\ell}(z,\bar{z}) = \sum_{n,m\geq 0} A_{n,m} k_{n+m +\beta+1}(z) k_{n-m + \alpha+1/2} (\bar{z})$, which is consistent with the Hogervorst's formula~\cite{hogervorst2016dimensional}, up to a scaling factor, thereby completing the proof of this conjecture for the 3D case (see Supplemental Material~\cite{SM} Sec~D).

\mysec{Applications to the Conformal Bootstrap}
The technique we introduced has various implications for conformal bootstrap. Below, we list a few potential applications. The first application is determining the OPE coefficients for a given correlation function. As an example, we consider either a 1D correlation function or the chiral part of a 2D correlation function \( f(z) \) with an integer-spaced spectrum. In such cases, we can expand
\(
f(z) = \sum_{n= 0}^\infty a_n z^{h+n}
\)
in terms of the Frobenius series. A common task is to express this function as an expansion in terms of the 1D conformal block 
\begin{equation}
f(z) = \sum_{n= 0}^\infty c_n k_{h+n}(z),\notag
\end{equation}
and determine the OPE coefficients \( c_n \). Several techniques have been developed for this purpose, such as the alpha space transform \cite{hogervorst2017crossing} and, more generally, the conformal inversion formula \cite{caron2017analyticity, simmons2018spacetime}. However, these approaches often involve complex integrations. In contrast, our technique only requires the application of fractional derivatives 
\(
(T_z^{1/2} f)(z),
\)
followed by applying the quadratic transformation (see Supplemental Material~\cite{SM} Sec.~B), which yields
\(
     c_n = \frac{4^h\Gamma(h-1/2)}{\Gamma(h)} \frac{(2h-1)_n (-1)^n}{n!} 
     \sum_{m=0}^n   \frac{(-n)_m (n+2h-1)_m }{(h)_m^2 } a_m,
\)
which provides a direct method for computing the operator product expansion (OPE) coefficients \( c_n \) from the power series coefficients \( a_n \). Note that there is a singularity at \( h = 0 \). However, this issue can be circumvented by subtracting the identity operator, which effectively shifts \( h \) by one.

Moreover, our approach enables the direct expansion of Virasoro blocks in the Generalized Minimal Model in terms of global blocks \cite{behan2018unitary}. These blocks satisfy the BPZ equations, which originate from the null condition of the finite-order differential operator \( \mathcal{L}_{BPZ} \). Interestingly, Ref.~\cite{behan2018unitary} demonstrates that, for the lowest-order case, the OPE coefficients obey a recurrence relation crucial for proving their positivity. However, whether such relations exist more generally remains an open question. To address this, we may consider the BPZ operator under the fractional transform 
\begin{equation}
\mathcal{L}_{BPZ} \rightarrow \tilde{\mathcal{L}}_{BPZ} \ldef \mathcal{T}^{1/2} \mathcal{L}_{BPZ}  \mathcal{T}^{-1/2}.\notag
\end{equation}
It is straightforward to observe that, in this new representation, the BPZ equations remain finite-order ordinary differential equations, thereby permitting Frobenius series solutions, $\sum_n c_n \rho^{n+s-1/2}$, from which the OPE coefficients \( c_n \) can be directly extracted. Furthermore, this suggests that \( c_n \) generally satisfies recurrence relations, a natural consequence of locality, as the differential equation remains of finite order.

\begin{figure}[!htb]
\centering
\includegraphics[width=1\linewidth]{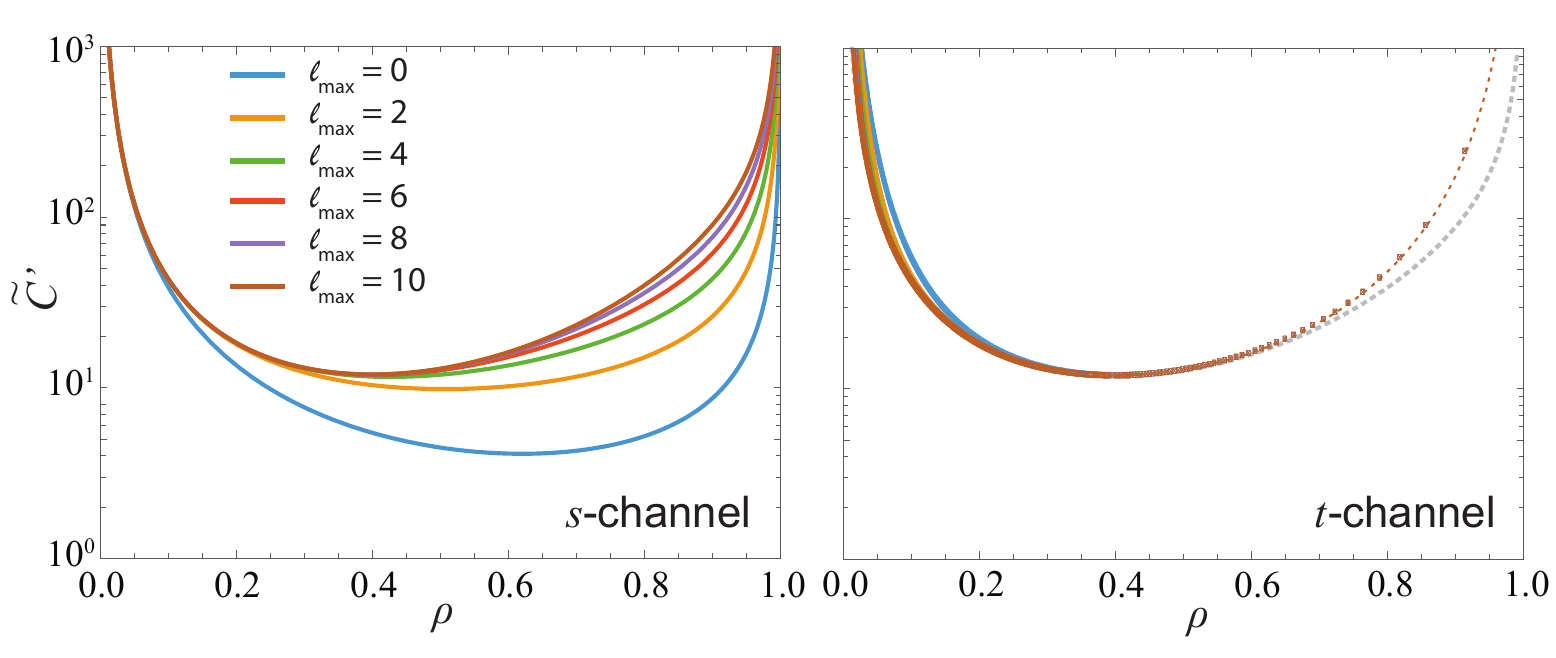}
\caption{Numerical test of the crossing symmetry relation Eq.~\eqref{eq:crossing}. The function $\tilde C'(\rho,\bar\rho)$ is evaluated along the diagonal $\rho=\bar\rho$ with spin truncations $\ell_\mathrm{max}=0,2,\ldots,10$. Results are shown for the $s$-channel (left) and $t$-channel (right), using OPE data from Ref.~\cite{simmons2017lightcone}. Rapid convergence is observed as $\ell_\mathrm{max}$ increases. The thick dashed brown curve shows the isolated identity contribution $C_0'(\rho)C_0'(\bar\rho)$, which dominates at large $\rho$, while the dashed gray curve reproduces the $s$-channel result with $\ell_\mathrm{max}=10$ for comparison.
}    
\label{fig:tildeC}
\end{figure}

Finally, our technique applies to the 3D conformal bootstrap. Instead of solving the crossing equation Eq.~\eqref{eq:crossing0} directly, we apply the $\mathcal{T}^{1/2}$ transformation, defining $\tilde C(\rho,\bar \rho) \ldef \mathcal{T}^{1/2} C(z,\bar z)$. The crossing symmetry can then be reformulated as
\begin{equation}
\tilde C \ldef \sum_{\dd,\ell} c_{\dd,\ell} \tilde G_{\dd,\ell}(\rho,\bar \rho)
= C_0(\rho)C_0(\bar\rho)+\sum_{\dd,\ell} c_{\dd,\ell}\, \mathcal{K}\,\tilde G_{\dd,\ell}(\rho,\bar \rho), \notag
\end{equation}
where the conformal block in our representation, $\tilde G_{\dd,\ell}(\rho,\bar \rho)$, is given by Eq.~\eqref{eq:G}, and $C_0(\rho)\ldef \frac{\Gamma(\dd_\phi-1/2)}{\Gamma(\dd_\phi)}\left(4\rho/(1-\rho)^2\right)^{\dd_\phi-1/2}$. The crossing-kernel operator $\mathcal{K}\ldef \mathcal{K}_z \mathcal{K}_{\bar z}$ maps $s$-channel blocks into $t$-channel blocks~\cite{hogervorst2017crossing,hogervorst2017crossing2,karateev2018weight,liu2019d,chen2019conformal}, with
\begin{equation}
    \mathcal{K}_x \ldef T_x^{1/2}\left(\frac{x}{1-x}\right)^{\Delta_\phi} T_{1-x}^{-1/2}, \notag
\end{equation}
whose explicit form is provided in Supplemental Material~\cite{SM} Sec.~E.  

Practically, it is convenient to use the derivative 
$\tilde C'(\rho,\bar \rho)\ldef \partial_{\rho,\bar \rho} \tilde C(\rho,\bar \rho)$, since the half-derivatives $T^{\pm 1/2} f$ in Eqs.~\eqref{eq:Thalf} involve derivatives of $f$. Equivalently, the crossing relation becomes
\begin{equation}
\sum_{\dd,\ell} c_{\dd,\ell}\, \tilde G_{\dd,\ell}'(\rho,\bar \rho)
= C_0'(\rho)C_0'(\bar\rho)
+ \sum_{\dd,\ell} c_{\dd,\ell}\, (\mathcal{K}\tilde G_{\dd,\ell})',
\label{eq:crossing}
\end{equation}

To test Eq.~\eqref{eq:crossing}, we performed a concrete 3D Ising example using the numerical data of ~\cite{simmons2017lightcone}, truncating the spin sum at $\ell_{\max}=0,2,4,\ldots,10$, and evaluated $C'(\rho,\bar\rho)$ in both $s$- and $t$-channels (Fig.~\ref{fig:tildeC}). We find rapid convergence as $\ell_{\max}$ increases. Moreover, the $s$- and $t$-channels agree very well, with deviations only at large $\rho$, where contributions from large-spin operators dominate. With our analytic 3D block formula, Eq.~\eqref{eq:crossing} thus provides a fully determined formulation of the 3D conformal bootstrap. Unlike existing approaches, where treating 3D blocks is analytically cumbersome, our method offers a promising route for simplifying the bootstrap equations. Finally, it is straightforward to convert back to the standard representation $C=\mathcal{T}^{-1/2}\tilde C$ using the inverse transform, as illustrated in Supplemental Material~\cite{SM} Sec.~F.

\mysec{Discussion}
Our results uncover an unexpected and conceptually broad connection between fractional calculus and conformal blocks, leading to the first exact closed-form solution for the 3D blocks with arbitrary spin. This achievement resolves Hogervorst’s conjecture in full generality for spinning operators, a long-standing open problem in the conformal bootstrap. Unlike in 2D and 4D, where compact analytic forms of conformal blocks have long been available, no such representation existed in 3D previously. Filling this gap is of particular importance, since 3D conformal field theories, most prominently the 3D Ising model, are central to the modern bootstrap program. Beyond resolving Hogervorst’s conjecture, our approach provides a novel analytical tool that both simplifies block computations and offers new leverage for analytic and numerical studies.  

A natural next step is to extend the method to mixed correlators, potentially generalizing Hogervorst’s formula to this more general setting. Our fractional derivative $T^{1/2}$ transforms the 1D block into a power function, greatly simplifying calculations. For non-identical operators, however, the 1D block takes the form $z^h \, {}_2F_1(h+a,h+b,2h,z)$, where $a \ldef \Delta_1-\Delta_2$ and $b \ldef \Delta_3-\Delta_4$, and in this case a direct application of $T^{1/2}$ does not yield a simple expression. Another open direction concerns dimensional generality: while we have focused on 3D, Hogervorst’s conjecture applies to arbitrary $d$. Addressing both challenges will require generalizing the integral transform in Eq.~\eqref{eq:T1} while preserving key algebraic properties, such as Eq.~\eqref{eq:xD}, to maintain the simplicity of the Casimir equation. Investigating such generalizations is an important avenue for future work.  

Despite these open directions, our findings show that fractional calculus provides both a powerful computational tool and a cleaner analytic structure for conformal blocks. In particular, our compact ${}_4F_3$ representation converges much more rapidly than traditional expansions and remains finite in regimes where standard blocks develop logarithmic singularities. These features offer a new analytic perspective on conformal bootstrap equations, with potential impact well beyond the identical-operator case studied here. More broadly, the emergence of a structural link between fractional calculus and conformal blocks was unexpected, and suggests that this framework may become valuable in a wider range of bootstrap applications. Finally, since fractional operators are widely used across physics and applied mathematics, this bridge may also open avenues for cross-fertilization beyond conformal field theory itself.

\begin{acknowledgments}
\textbf{Acknowledgments.} We thank Connor Behan, Agnese Bissi, Slava Rychkov, and Aninda Sinha for insightful discussions and helpful suggestions that improved this work.
\end{acknowledgments}

\appendix
\section{Fractional Derivatives}

In this section, we specify the fractional derivatives $T^{\pm}$ used in our work. We begin with the general definition
\begin{align}
(T^\delta_x f)(x) &= \frac{x^{1-\delta}}{\Gamma(\delta)} \int_0^x f(t)\,(x-t)^{\delta-1}\, t^{-\delta-1}\, dt \\
&= \frac{x^{-\delta}}{\Gamma(\delta)} \int_0^1 f(xt)\,(1-t)^{\delta-1}\, t^{-\delta-1}\, dt. \notag
\end{align}

Suppose $f(t) \sim t^h$ as $t \to 0$. For $\delta = \tfrac{1}{2}$ we obtain
\begin{align}
(T^{1/2}_x f)(x) &= \frac{x^{1/2}}{\sqrt{\pi}} \int_0^x f(t)\,(x-t)^{-1/2}\, t^{-3/2}\, dt \\
&= \frac{x^{-1/2}}{\sqrt{\pi}} \int_0^1 f(xt)\,(1-t)^{-1/2}\, t^{-3/2}\, dt, \notag
\end{align}
which converges for $h > 1/2$.

Formally, the inverse operation reads
\begin{align}\label{eq:Tinv}
(T^{-1/2}_x f)(x) &= -\frac{x^{3/2}}{2\sqrt{\pi}} \int_0^x f(t)\,(x-t)^{-3/2}\, t^{-1/2}\, dt \\
&= -\frac{x^{1/2}}{2\sqrt{\pi}} \int_0^1 f(xt)\,(1-t)^{-3/2}\, t^{-1/2}\, dt. \notag
\end{align}
However, this integral diverges due to the singular behavior as $t \to 1$. To obtain an analytic continuation, we perform an integration by parts, yielding
\begin{align}\label{eq:TinvAC}
(T^{-1/2}_x f)(x) &= \frac{x^{1/2}}{\sqrt{\pi}} \int_0^x f'(t)\,(x-t)^{-1/2}\, t^{1/2}\, dt \\
&= \frac{x^{3/2}}{\sqrt{\pi}} \int_0^1 f'(xt)\,(1-t)^{-1/2}\, t^{1/2}\, dt, \notag
\end{align}
which converges for $h > -1/2$.

Since
\begin{equation}
    T_x^{-1/2} f = T_x^{-1} T_x^{1/2} f 
    = -\frac{d}{d (1/x)} T_x^{1/2} f 
    = x^2 \big(T_x^{1/2} f\big)',
\end{equation}
we find the useful relation
\begin{align}\label{eq:Thalf2}
(T^{1/2}_x f)(x)' &= \frac{x^{-3/2}}{\sqrt{\pi}} \int_0^x f'(t)\,(x-t)^{-1/2}\, t^{1/2}\, dt \\
&= \frac{x^{-1/2}}{\sqrt{\pi}} \frac{d}{dx} \int_0^x f(t)\,(x-t)^{-1/2}\, t^{-1/2}\, dt \notag \\
&= \frac{x^{-1/2}}{\sqrt{\pi}} \int_0^1 f'(xt)\,(1-t)^{-1/2}\, t^{1/2}\, dt. \notag
\end{align}

Finally, introducing the radial coordinate $z = \tfrac{4\rho}{(1+\rho)^2}$ and defining $F(\rho) := f(z)$, one finds
\begin{align}\label{eq:Tinvrho}
(T^{-1/2}_z F)(\rho) 
= \frac{2\rho^{1/2}}{\sqrt{\pi}} \int_0^\rho 
F'(\rho') \sqrt{\frac{\rho'}{(\rho-\rho')(1-\rho\rho')}}\, d\rho'.
\end{align}

\section{Quadratic Transformation}\label{sec:qt}

In this section, we derive a general formula for the quadratic transformation, which will be used extensively in our subsequent calculations. This transformation plays a central role in simplifying hypergeometric functions and in connecting different representations of conformal blocks.

For a Frobenius series $f(x) = \sum_{n\geq 0} a_n x^{s+n}$, a quadratic transformation can be expressed as
\begin{equation}\label{eq:qt}
\left(\frac{4x}{(1+x)^2} \right)^s f\left(\frac{4x}{(1+x)^2}\right) = (4x)^s \sum_{k=0}^\infty  c_k x^k, 
\end{equation}
where the coefficient 
\begin{equation} \label{eq:coeff0}
    c_k = \frac{(2s)_k (-1)^k}{k!} \sum_{n=0}^k a_n \frac{(-k)_n (k+2s)_n }{(s)_n (s+1/2)_n }. 
\end{equation}
The proof unfolds straightforwardly:
\begin{equation}
\begin{split}
(1+x)^{-2s}&f\left(\frac{4x}{(1+x)^2}\right) = \sum_{n\geq 0}a_n(4x)^n (1+x)^{-2(n+s)} \\
&=  \sum_{n\geq 0} a_n (4x)^n \sum_{m \geq 0} \frac{(-2(n+s))!}{(-2n-2s-m)!m!} x^{m} \\
&= \sum_{k\geq 0}  x^{k} \sum_{n\geq 0}^{k}  a_n \frac{4^n(-2(n+s))!}{(-n-2s-k)!(k-n)!}  \\
&=  \sum_{k\geq 0}  \frac{(2s)_k}{k!} (-x)^k \sum_{n\geq 0}^{k} a_n \frac{(-k)_n (k+2s)_n }{(s)_n (s+1/2)_n }.
\end{split} 
\end{equation}
As a result, we obtain the following quadratic transformation for the hypergeometric function:
\begin{align}\label{eq:qt0}
&\left(\frac{4x}{(1+x)^2} \right)^s \pFq{p}{q}{a_1, \ldots, a_p}{b_1, \ldots, b_q}{\frac{4x}{(1+x)^2}} \notag \\
&= (4x)^s \sum_{k=0}^\infty \frac{(2s)_k}{k!} \pFq{p+2}{q+2}{-k, k + 2s, a_1, \ldots, a_p}{s, s+1/2, b_1, \ldots, b_q}{1} (-x)^k.
\end{align}
Focusing on the Gauss hypergeometric function ${}_2 F_1$ with $a_1 = s$ and $a_2 = s+1/2$, the coefficient ${}_{p+2} F_{q+2}$ simplifies to ${}_2 F_1(-k, k+2s, b; 1) = \frac{\Gamma(b)\Gamma(b-2s)}{\Gamma(b-k-2s)\Gamma(b+k)} = \frac{(2s - b + 1)_k}{(b)_k} (-1)^k $, which reiterates the standard quadratic transformation:
\begin{align}\label{eq:qt1}
\left(\frac{4x}{(1+x)^2} \right)^s\pFq{2}{1}{s, s+1/2}{b}{\frac{4x}{(1+x)^2}} 
\\\notag =(4x)^s \pFq{2}{1}{2s, 2s-b+1}{b}{x}. 
\end{align}
By setting $s = h-1/2$ and $b = 2s+1 = 2h$ in Eq. \eqref{eq:qt1}, we obtain 
\begin{equation} \label{eq:qt2}
(z/4)^{h-1/2}\pFq{2}{1}{h, h-1/2}{2h}{z} =\x ^{h-1/2}.
\end{equation}

Lastly, we compute the quadratic transformation for $ F(\a,\b; x) = 4^{-s}\sum_{n\geq 0} a_n x^{s+n}$, where
\begin{equation}
    a_n = \frac{(s)_n(s+1/2)_n}{(2s+1/2)_n n!} \pFq{3}{2}{-n, 1/2, 1/2}{1+\a,1+\b}{1},
\end{equation}
with $s = (\alpha+\beta+1/2)/2$. The coefficient in Eq. \eqref{eq:coeff0} is given by
\begin{equation}
\begin{split}
    c_k 
    & = \frac{(2s)_k (-1)^k}{k!} \sum_{n=0}^k \frac{(-k)_n (k+2s)_n}{(2s+1/2)_n n!} \pFq{3}{2}{-n, 1/2, 1/2}{1+\a,1+\b}{1}   \\
    & = \frac{(2s)_k (-1)^k}{k!} \sum_{n=0}^k \frac{(-k)_n (k+2s)_n}{(2s+1/2)_n n!} \sum_{m=0}^n \frac{(-n)_m (1/2)_m^2}{m! (1+\a)_m (1+\b)_m}  \\
    & = \frac{(2s)_k}{k!} \frac{(1/2)_k (\a+1/2)_k (\b+1/2)_k}{(\a+1)_k(\b+1)_k (2s+1/2)_k}.
\end{split}
\end{equation}

\section{Coefficients of Recursive Relationship} 

Recall the recursive relationship of coefficients,
\begin{align}
f_1A_{n,m} - f_2 A_{n-1,m} - f_3 A_{n,m-1} + f_4 A_{n-1,m-1} = 0,\notag
\end{align}
where $f_1 \ldef ( n-m + \a)(n + m + \b+\nu-1/2) \left( (n + m + \b)^2 + (n-m + \a)^2 - \a^2-\b^2\right)$, $f_2 \ldef (n-m + \a-1/2) (n + m + \b+\nu-1) \left( (n + m + \b+\nu-1)^2 + (n-m + \a + \nu-1)^2 - \a^2-\b^2 \right) $, $f_3 \ldef (n-m + \a+ 1/2 )  (n+m+\b+\nu-1) \left( (n + m + \b+\nu-1)^2 + (n-m + \a+1-\nu)^2 - \a^2-\b^2\right) $, and $f_4 \ldef ( n-m + \a) ( n + m + \b+\nu -3/2)  \left( (n + m + \b+2(\nu-1))^2 + (n-m + \a)^2 - \a^2-\b^2 \right)$.

\section{Identification with Hogervorst's Formula}

In this section, we demonstrate that our result is identical to the 3D decomposition formula of Hogervorst~\cite{hogervorst2016dimensional}, up to an overall normalization factor. The formula reads
\begin{equation}
G^{(3d)}_{\Delta,\ell}(z,\bar z) \;=\; \sum_{n=0}^{\infty} \sum_{j} 
\mathcal A_{n,j}(\Delta,\ell)\, G^{(2d)}_{\Delta+2n,j}(z,\bar z),
\end{equation}
for $j = \ell, \ell-2, \ldots$. It is convenient to reparameterize $j = \ell - 2m$, with $m = 0,1,\ldots,\lfloor \ell/2 \rfloor$.  

The two-dimensional conformal block is given by
\begin{equation}
G^{(2d)}_{\Delta,\ell}(z,\bar z) 
\ldef \tfrac{1}{2}\,\Big[ k_{\Delta+\ell}(z)\, k_{\Delta-\ell}(\bar{z}) \;+\; (z \leftrightarrow \bar{z}) \Big],
\end{equation}
with
\begin{equation}
k_h(z) \;\ldef\; z^h \, {}_2F_1(h,h,2h;z).
\end{equation}

The coefficient $\mathcal A_{n,j}$ was conjectured to take the form
\begin{align}
\mathcal A_{n,j}(\Delta,\ell) &= C(\Delta,\ell)\, Z_\ell^j \,
\frac{\left( \tfrac{\Delta+j}{2} \right)_n \left( \tfrac{\tau+\ell-j+1}{2} \right)_n}
     {\left( \tfrac{\Delta+j-1}{2} \right)_n \left( \tfrac{\tau+\ell-j}{2} \right)_n}
\frac{(1/2)_n \, (\Delta-1)_{2n} \, (\tfrac{\Delta+\ell}{2})_n \, (\tfrac{\tau}{2})_n}
     {16^n \, n! \, (\Delta-\nu)_n \, (\Delta-\nu-1/2+n)_n \, (\tfrac{\Delta+\ell+1}{2})_n \, (\tfrac{\tau+1}{2})_n},
\end{align}
where $C(\Delta,\ell)$ is an overall normalization constant, $\tau \ldef \Delta-(\ell+d-2)$, $\nu = 1/2$, and 
\begin{equation}
Z_\ell^j = 2 \, \frac{(1/2)_m (1/2)_{\ell-m}}{m!(\ell -m)!}, 
\qquad (d=3).
\end{equation}

Using our notation $\a \ldef (\Delta+\ell-1)/2$ and $\b \ldef (\Delta-\ell-2)/2$, the coefficient can be expressed as
\begin{widetext}
\begin{align}
\mathcal A_{n,j}(\Delta,\ell) 
&= 2 \cdot 4^{-2n}\,
\frac{(1/2)_n \, (1/2+\a)_n \, (\b+1/2)_n \, (1/2+\a+\b)_n}{n! \,(1+\a)_n (1+\b)_n (1+\a+\b)_n } \times \frac{(1/2)_m \, (1/2+\a-m)_n (1+\b+m)_n (1/2)_{\ell-m}}
{m!\, (\a-m)_n (1/2+\b+m)_n (\ell -m)!}.
\label{eq:A}
\end{align}
\end{widetext}

Now recall the solution in the main text,
\begin{equation}\label{eq:main}
G_{\Delta,\ell}(z,\bar z) \;=\; \sum_{n,m\geq 0} 
A_{n,m} \, \tilde k_{n+m+\b+1}(z)\,\tilde k_{n-m+\a+1/2} (\bar z),
\end{equation}
where $\tilde k_h(z) \ldef c_h\, k_h(z)$ includes a normalization prefactor
\begin{equation}
c_h \;\ldef\; 4^{1/2-h}\, \frac{\Gamma(h)}{\Gamma(h-1/2)},
\end{equation}
and the coefficient $A_{n,m}$ is given by
\begin{widetext}
\begin{align}
A_{n,m} &= \frac{(1/2)_n (1/2+\a)_n (\b+1/2)_n (1/2+\a+\b)_n}{n!(1+\a)_n (1+\b)_n (1+\a+\b)_n }\times \frac{(1/2)_m (1/2-\a)_m (\b+1/2)_m (1/2-\a+\b)_m}{m!(1-\a)_m (1+\b)_m (1-\a+\b)_m }.
\end{align}
\end{widetext}
This satisfies the symmetry $A_{n,m} = A_{n,\ell-m}$ for $m = 0,1,\ldots,\ell$, which naturally generates the exchange term $(z \leftrightarrow \bar z)$ in the 2D block.  

The extra normalization factor gives rise to
\begin{align}
c_{n+m+\b+1}\, c_{n-m+\a+1/2} 
= c_{\a+1/2}\, c_{\b+1}\, \times \notag \\  4^{-2n}\,
\frac{(\a+1/2)_{n-m}(\b+1)_{n+m}}{(\a)_{n-m}(\b+1/2)_{n+m}}.  \notag
\end{align}

Using the identities 
\[
(x)_{n+m} = (x+m)_n (x)_m,\qquad
(1-x)_m (x)_{n-m} = (-1)^m (x-m)_n,\qquad
(x)_{n-m}(1-x-n)_m = (-1)^m (x)_n,
\]
together with $\ell = \a - \b - 1/2$, one finds
\begin{align}
    \frac{(1/2 -\a)_m}{(1-\a)_m} \cdot \frac{(\a+1/2)_{n-m}}{(\a)_{n-m}} &= \frac{(\a+1/2-m)_{n}}{(\a-m)_{n}}, \notag \\
    \frac{(\b+1/2)_m}{(1+\b)_m} \cdot \frac{(\b+1)_{n+m}}{(\b+1/2)_{n+m}} &= \frac{(1+\b+m)_{n}}{(1/2+\b+m)_{n}}, \notag \\
    \frac{(1/2 -\a+\b)_m}{(1-\a+\b)_m } &= \frac{(-\ell)_m}{(1/2-\ell)_m } =  \frac{\ell !}{(1/2)_\ell } \cdot \frac{(1/2)_{\ell-m}}{(\ell-m)!}. \notag
\end{align}

Comparing with Eq.~\eqref{eq:A}, we conclude the identity
\begin{equation}
    \mathcal A_{n,j}(\Delta,\ell) \;=\; A_{n,m}\, c_{n+m+\b+1}\, c_{n-m+\a+1/2},
\end{equation}
with overall normalization
\begin{equation}
    C(\Delta,\ell) = c_{\a+1/2} \, c_{\b+1} \,\frac{\ell !}{(1/2)_\ell }  
    = \frac{\ell !}{(1/2)_\ell } 
    \frac{4^{1-\Delta}\, \Gamma\!\left(\tfrac{\Delta+\ell}{2}\right)\Gamma\!\left(\tfrac{\Delta-\ell}{2}\right)}
         {\Gamma\!\left(\tfrac{\Delta+\ell-1}{2}\right)\Gamma\!\left(\tfrac{\Delta-\ell-1}{2}\right)}.
\end{equation}

\section{Crossing Kernel}

We now turn to the analysis of the crossing kernel. Define
\begin{equation}
  K_z \;=\;  T_z^{1/2} \,\Big(\tfrac{z}{1-z}\Big)^{\Delta_\phi} \, T_{1-z}^{-1/2},   
\end{equation}
which satisfies the involution property
\begin{equation}
    K_z K_{1-z} \;=\; K_{1-z} K_z \;=\; 1.
\end{equation}

Using Eq.~\eqref{eq:Thalf2}, we find
\begin{align}
    (K_{z} f)(z)' = \frac{z^{-1/2}}{\sqrt\pi}\,\frac{d}{dz}
    \int_{1-z}^1 \frac{g(t)}{\sqrt{t-(1-z)}}\, dt 
    \end{align}
where
\begin{equation}
    g(t) \;\ldef\; (1-t)^{-1/2+\Delta_\phi}\, t^{-\Delta_\phi}\, \big(T_z^{-1/2} f\big)(t).
\end{equation}

Switching to radial coordinates with $t \mapsto \rho'$, we define
\begin{equation}
    G(\rho') \;\ldef\; g(t) \;=\; (1+\rho')\,(1-\rho')^{-1+2\Delta_\phi}\,(4 \rho')^{-\Delta_\phi}\, \big(T_z^{-1/2}F\big)(\rho'),
\end{equation}
where $F(\rho) := f(4\rho/(1+\rho)^2)$, so that
\begin{align}
    &\int_{1-z}^1 \frac{g(t)}{\sqrt{t-(1-z)}}\, dt 
    \notag\\&= 2 (1+\rho_t) \int_{\rho_t}^1 
    \frac{G(\rho')}{\sqrt{(\rho'-\rho_t)(1-\rho'\rho_t)}} 
    \frac{1-\rho'}{(1+\rho')^2}\, d\rho'. \notag
\end{align}

Thus, the crossing kernel action takes the form
\begin{widetext}
\begin{align}
    \frac{d}{dz}\,(K_{z} F)(\rho_t) 
    = -\frac{(1+\rho_t)^4}{2\sqrt\pi\,(1-\rho_t)^2}\,
    \frac{d}{d\rho_t}\Bigg[
    (1+\rho_t) \int_{\rho_t}^1 \frac{(1-\rho')^{2\Delta_\phi}\,(4 \rho')^{-\Delta_\phi}\, (T_z^{-1/2}F)(\rho')}
    {(1+\rho')\sqrt{(\rho'-\rho_t)(1-\rho'\rho_t)}}\, d\rho'\Bigg].
\end{align}
\end{widetext}

\section{Numerical Results}

To validate our analytic ${}_4F_3$ representation, we performed several numerical checks against known results.  

\begin{figure}[!htb]
\centering
\includegraphics[width=0.8\linewidth]{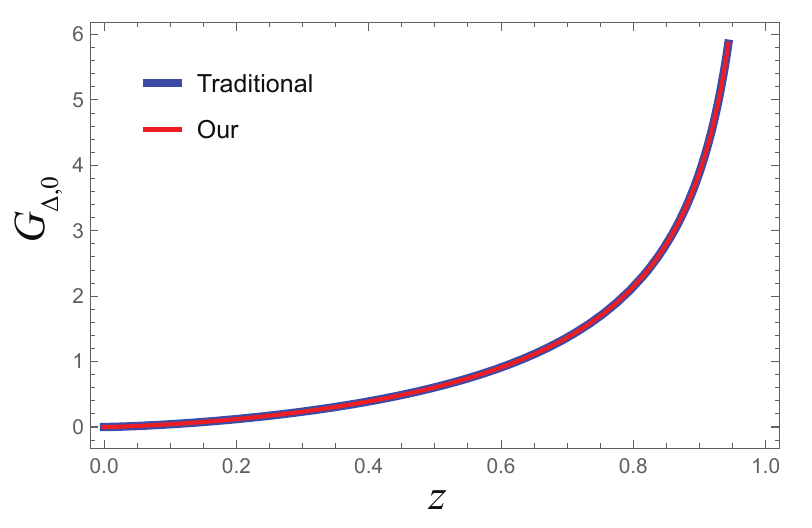}
\caption{Direct numerical comparison of the scalar conformal block along the diagonal $z=\bar z$. Our ${}_4F_3$ formula (solid line) agrees with the standard numerical evaluation within machine precision.}    
\label{fig:compare}
\end{figure}

Figure~\ref{fig:compare} shows the comparison between the scalar block computed using our closed-form expression and the well-established numerical methods. The two results coincide to within numerical precision, confirming the correctness of our formula.  

\begin{figure}
\centering
\includegraphics[width=1\linewidth]{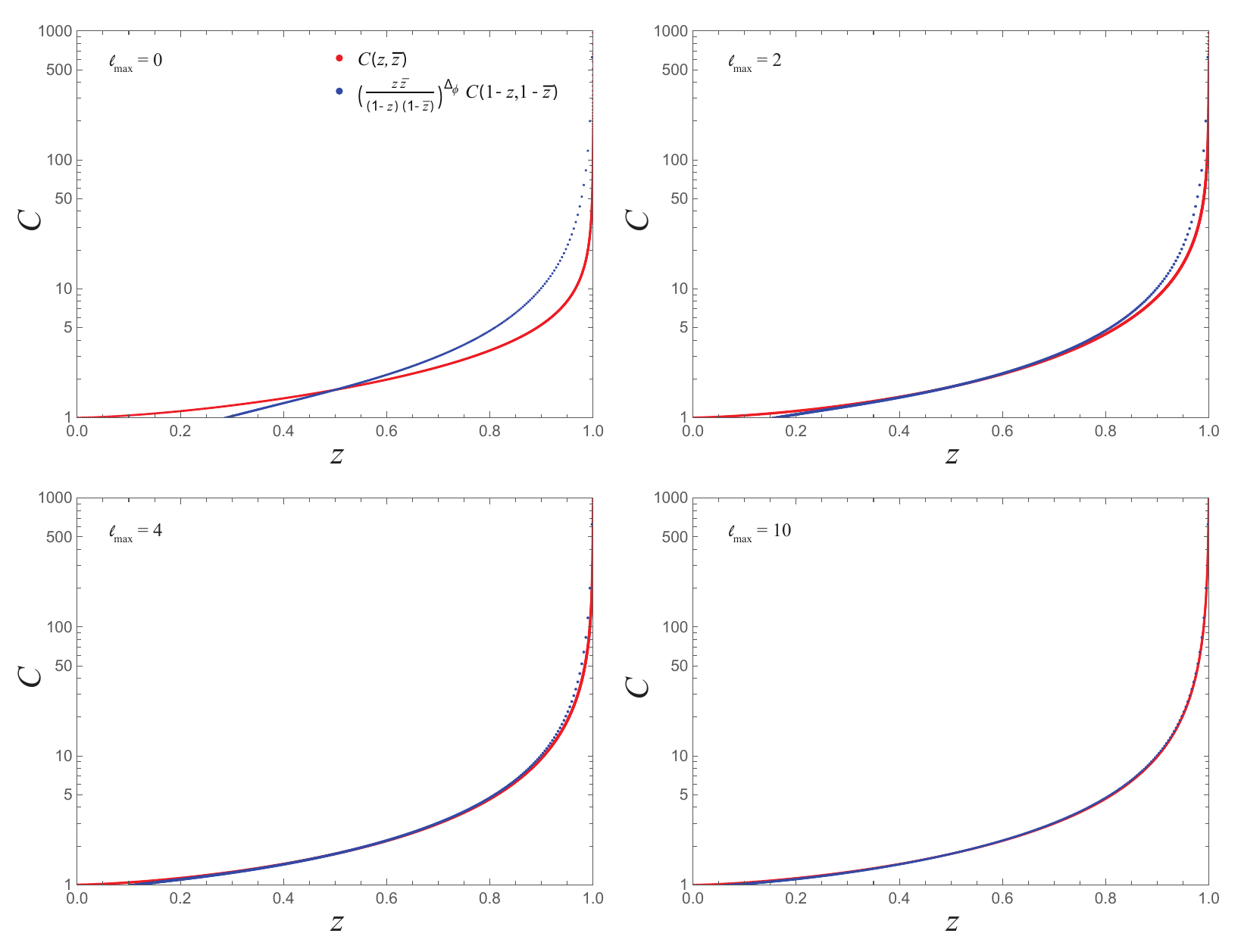}
\caption{Numerical test of crossing symmetry for the four-point correlator, obtained using the inverse transform of our ${}_4F_3$ block representation. The sum is truncated at spins $\ell_{\max}=0,2,4,\ldots,10$. The OPE data are taken from Ref.~\cite{simmons2017lightcone}.}    
\label{fig:C}
\end{figure}

In Fig.~\ref{fig:C}, we test crossing symmetry using OPE data from the 3D Ising model~\cite{simmons2017lightcone}. The correlator is first constructed in our representation as $\tilde C(\rho,\bar\rho)$ and then mapped back to the standard form $C(z,\bar z)$ via the inverse transform $C=\mathcal{T}^{-1/2}\tilde C$. The sum is truncated at spins $\ell_{\max}=0,2,\ldots,10$. We find that the $s$- and $t$-channel results converge rapidly with increasing $\ell_{\max}$ and agree to high precision, providing a strong validation of both our analytic block representation and its practical utility in bootstrap calculations.

\end{document}